\documentclass[a4paper]{jpconf}

\usepackage[T1]{fontenc}
\usepackage[utf8]{inputenc}
\usepackage[english]{babel}
\usepackage[numbers,square,sort&compress]{natbib}
\usepackage{graphicx}
\usepackage[usenames,dvipsnames]{xcolor}
\usepackage{mathtools}
\usepackage{dsfont}
\usepackage{lmodern}
\usepackage{siunitx}
\usepackage[final]{microtype}

\newcommand*{\+}{\hspace*{0.08335em}}
\newcommand*{\?}{\hspace*{-0.08335em}}
\newcommand*{\beq}{\begin{equation}}
\newcommand*{\eeq}{\end{equation}}
\newcommand*{\Nf}{N_{\textup{f}}}
\newcommand*{\Nc}{N_{\textup{c}}}
\newcommand*{\dd}{\textup{d}}
\newcommand*{\upB}{\textup{B}}
\newcommand*{\upu}{\textup{u}}
\newcommand*{\upd}{\textup{d}}
\newcommand*{\ups}{\textup{s}}
\newcommand*{\upq}{\textup{q}}
\newcommand*{\muq}{\mu_{\upq}}
\newcommand*{\muB}{\mu_{\upB}}

\DeclarePairedDelimiterX{\expval}[1]{\langle}{\rangle}{#1}
\DeclareSIUnit{\eV}{\electronvolt}

\begin{document}

\title{Mesons at finite chemical potential and the Silver-Blaze property of QCD}

\author{Pascal J. Gunkel, Christian S. Fischer and Philipp Isserstedt}

\address{Institut f\"ur Theoretische Physik, Justus-Liebig-Universit\"at Gie\ss{}en, 35392 Gie\ss{}en, Germany}

\ead{Pascal.Gunkel@physik.uni-giessen.de}

\begin{abstract}
We summarize our results for light (pseudo-)scalar mesons at finite chemical potential and vanishing temperature. We extract the meson bound state wave functions, masses, and decay constants up to the first-order phase transition from the homogeneous Bethe-Salpeter equation and confirm the validity of the Silver-Blaze property. For this purpose, we solve a coupled set of truncated Dyson-Schwinger equations for the quark and gluon propagators of QCD in Landau gauge.
\end{abstract}

\section{\label{sec:intro}Introduction}

The QCD phase diagram and the properties of hadrons in medium are strongly connected. 
The properties of pseudoscalar and scalar meson degrees of freedom are tied to dynamical 
chiral symmetry breaking and its restoration and play an important role in the vicinity 
of second-order phase transitions: they account for the effective long-range degrees of 
freedom and control the universal behavior.

At zero temperature and small chemical potential, QCD displays the so-called Silver-Blaze 
property: observables are unaffected by chemical potential as long as the baryon chemical 
potential $\muB$ is not large enough to create physical excitations. While this property
can be shown analytically for the case of finite isospin chemical potential, it is also 
extremely plausible for the case of finite baryon chemical potential
\cite{Cohen:1991nk,Cohen:2003kd,Cohen:2004qp} and has been demonstrated for heavy quark masses
in the lattice formulation of  Ref.~\cite{Fromm:2012eb}. At physical quark masses, lattice QCD is hampered by the fermion sign problem and one has to resort to extrapolations,
functional methods, and/or effective models.

In this work we summarize results presented in Ref.~\cite{Gunkel:2019xnh}. We approach QCD with the
non-perturbative functional method of Dyson-Schwinger equations (DSEs) and extend previous bound state
calculations for pion and sigma mesons to finite chemical potential and vanishing temperature. To this
end, we use a variant of a well-studied truncation scheme that includes the backcoupling of the quark
onto the gluon, see Ref.~\cite{Fischer:2018sdj} for a comprehensive review. We extract meson properties
from their Bethe-Salpeter equations (BSEs) and investigate the validity of the Silver-Blaze property.

\section{\label{sec:theoretical_framework}Theoretical framework}

\subsection{\label{sec:quark_gluon_propagators}Quark and gluon propagators}

\begin{figure}[t]
\begin{minipage}[t]{0.49\textwidth}%
\centering%
\includegraphics[scale=1,trim=16pt -40pt 0 0]{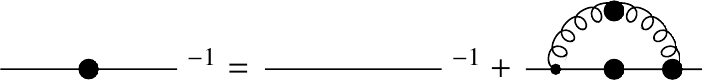}\\[-2em]%
\includegraphics[scale=1]{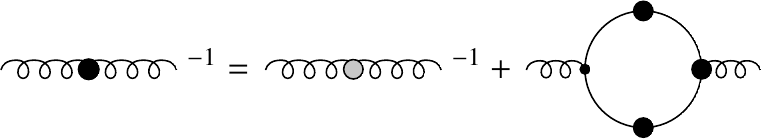}\\[1em]%
\includegraphics[width=0.9\textwidth]{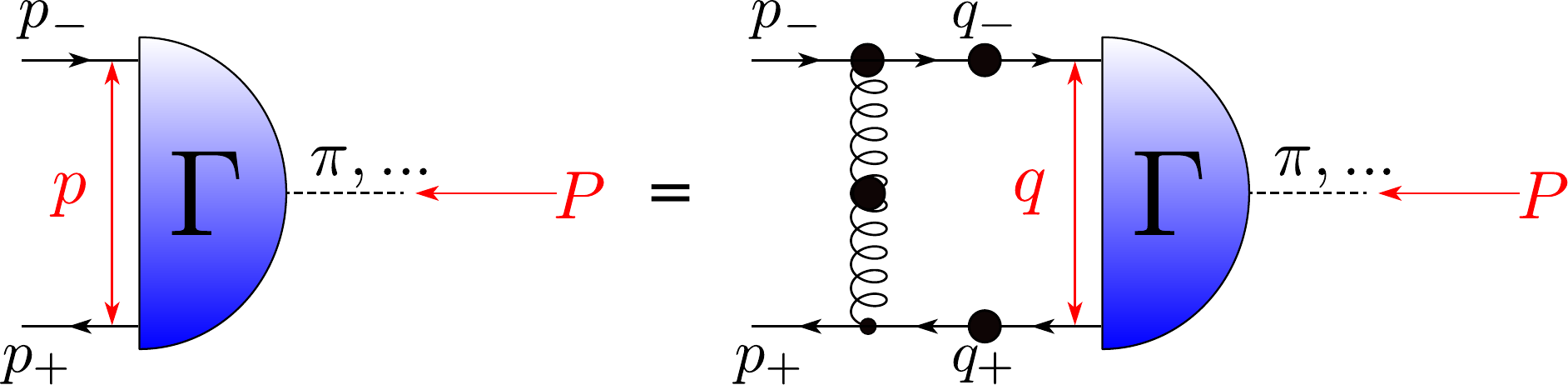}%
\end{minipage}%
\hfill%
\begin{minipage}[t]{0.49\textwidth}%
\flushright%
\includegraphics[width=0.9\textwidth,trim=0 100pt 0 0\textwidth]{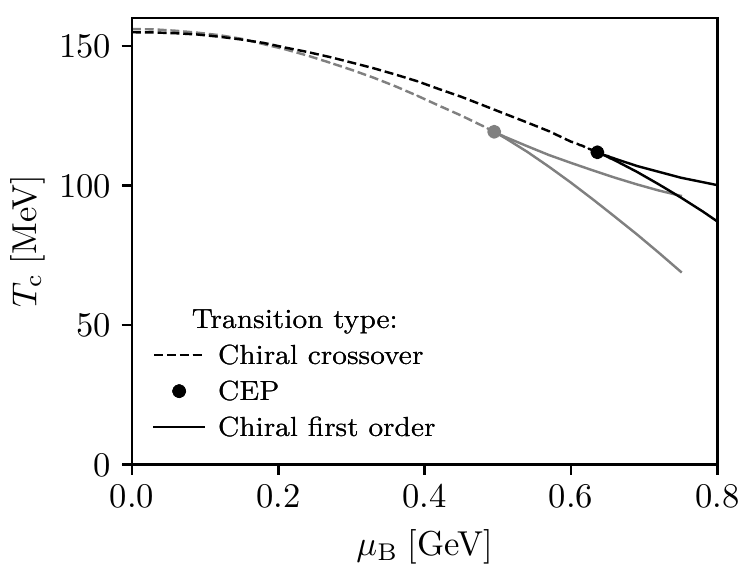}%
\end{minipage}%
\vspace*{+3mm}%
\caption{\label{fig:equations}%
Left: Truncated DSEs for the inverse quark (top) and gluon (center) propagators and graphical representation of the homogeneous BSE (bottom). Quark (gluon) propagators are denoted by solid (curly) lines. Dressed quantities are indicated by big dots. Right: Results for the QCD phase diagram from previous works 
\cite{Fischer:2018sdj,Fischer:2014ata,Isserstedt:2019pgx} (grey) and using the simplified truncation introduced in Ref.~\cite{Gunkel:2019xnh} (black). The DSE diagrams were drawn with JaxoDraw \cite{Binosi:2008ig}.}
\end{figure}

A necessary input for the meson BSE is the dressed quark propagator $S_f$ in the complex momentum plane. We calculate this propagator by solving the truncated DSEs for the quark and gluon propagators which are shown diagrammatically on the left side of Fig.~\ref{fig:equations}.

The truncation employed in this work is detailed in Ref.~\cite{Gunkel:2019xnh}. In the gluon DSE, we use a
temperature-dependent fit to quenched lattice data \cite{Fischer:2010fx,Maas:2011ez} for all diagrams already
present in Yang-Mills theory (subsummed by the diagram with the grey dot in Fig.~\ref{fig:equations}). The
remaining quark loop diagram couples the matter sector of QCD back onto the gluon and includes an implicit sum
over the considered $\Nf=2+1$ quark flavors $f\in\{\upu,\upd,\ups\}$. As in previous works
\cite{Fischer:2014ata,Isserstedt:2019pgx,Fischer:2012vc,Eichmann:2015kfa} we use an ansatz for the
quark-gluon vertex that contains an effective running coupling thereby providing a continuous transition
between the strongly interacting low-momentum region and the weakly interacting high-momentum perturbative
region. However, due to the numerical complexity induced by the additional study of mesons, we 
simplify the treatment introducing two further approximations: we neglect the chemical-potential dependence of the gluon propagator and strip the quark-gluon vertex in the quark DSE from terms motived by a Slavnov-Taylor identity, see
Ref.~\cite{Gunkel:2019xnh} for details. As a result, we obtain a rainbow-ladder like truncation which
greatly simplifies the associated interaction kernel in the BSE, cf.~below. The effect of these additional
approximations on the phase diagram can be assessed from the right diagram in Fig.~\ref{fig:equations}.
Compared to previous results \cite{Fischer:2018sdj,Fischer:2014ata,Isserstedt:2019pgx} the critical end point
is shifted to even larger chemical potential; correspondingly the curvature of the transition line is slightly
smaller. Overall, however, the effects on the critical end point are not too drastic and consequently the
additional approximations are justified.

\subsection{\label{sec:bsa}Bethe-Salpeter amplitudes}

In the DSE approach, we use the homogeneous BSE displayed on the left side of Fig.~\ref{fig:equations} to
calculate the wave function of mesonic bound states. We consider the Bethe-Salpeter amplitude (BSA)
$\Gamma_{X}(p,P)$ for pseudoscalar ($X = \textup{P}$) and scalar mesons ($X=\textup{S}$). They depend on the time-like total meson momentum $P$ and on the relative momentum $p$ between the quark and the antiquark.
Furthermore, the BSA depends on the dressed gluon propagator, the quark-gluon vertex, and the dressed quark propagator evaluated at complex momenta $q_{\pm}= q \pm P / \+2$ with $q=(\vec{q},q_4+i\muq)$ where $\muq$ denotes the quark chemical potential. The quark-gluon vertex $\Gamma_{\nu}^f(p,q,k)$ in the BSE
equals the one employed in the quark DSE and satisfies the axial-vector Ward-Takahashi identity. This is necessary to maintain the (pseudo-)Goldstone properties of the pseudoscalar mesons. Guided from results in vacuum
(see, e.g., Ref.~\cite{Fischer:2008sp}) we represent the scalar and pseudoscalar mesons by their
two leading Dirac tensor structures, viz.
\begin{align}
\Gamma_{\textup{P}}(p,P)=\;\gamma_{5}&\left\{E_{\textup{P}}(p,P)-i\gamma_{4}P_4 \+ F_{\textup{P}}(p,P)\right\}, \\
\Gamma_{\textup{S}}(p,P)=\,\,\;\mathds{1}&\left\{E_{\textup{S}}(p,P)-i\gamma_{4}P_4 \+ P\cdotp p\+F_{\textup{S}}(p,P)\right\} .
\label{eq:BSA_finite_mu}
\end{align}

\subsection{\label{sec:meson_properties}Meson properties}

As meson properties we study the meson mass and decay constants. The mass $m_X$ is directly extracted from the BSE, while the decay constant $f_X$ in vacuum is calculated via
\begin{align}
f_{X}P_{\mu}&=Z_{2}^f \+ \Nc \+ \int \frac{\dd^4 q}{(2\pi)^4} \Tr \bigl[ \+ j^{X}_{\mu} S_f(q_{+}) \+ \hat{\Gamma}_{X}(q,P) \+ S_f(q_{-}) \bigr] \, , \\
j^{X}_{\mu}&=\begin{cases}
             \gamma_{5}\gamma_{\mu} & \textup{for pseudoscalar mesons}\,, \\
             \gamma_{\mu} & \textup{for scalar mesons}\,,
            \end{cases}
\label{eq:meson_decay constant_vacuum}
\end{align}
with $\hat{\Gamma}_{X}$ being the normalized BSA using the Nakanishi method \cite{Nakanishi:1965zza} and $\Nc=3$ the number of colors. If we introduce chemical potential, the decay constant splits up into two parts \cite{Son:2001ff,Son:2002ci}:
\beq
f_{X}P_{\mu} \, \xrightarrow{\,\muq \neq 0\,} \, \left[f_{X}^{\textup{s}} \+ T_{\mu\nu}(v) + f_{X}^{\textup{t}}L_{\mu\nu}(v)\right] \? P_{\nu}
\label{eq:decay_constant_splitting_finite_mu}
\eeq
with the spatial and temporal meson decay constants $f_{X}^{\textup{s}}$ and $f_{X}^{\textup{t}}$
transversal and longitudinal to the assigned direction of the medium $v=(\vec{0},1)$, respectively.

\section{\label{sec:results}Results}

\subsection{\label{sec:quark_condensate_results}Quark condensate}

\begin{figure}[t]
\centering%
\includegraphics[width=0.47\textwidth]{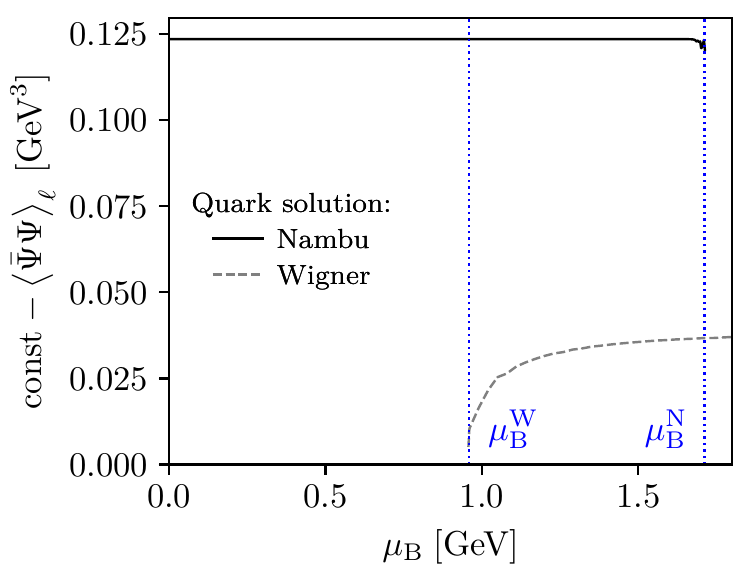}%
\hfill%
\includegraphics[width=0.47\textwidth]{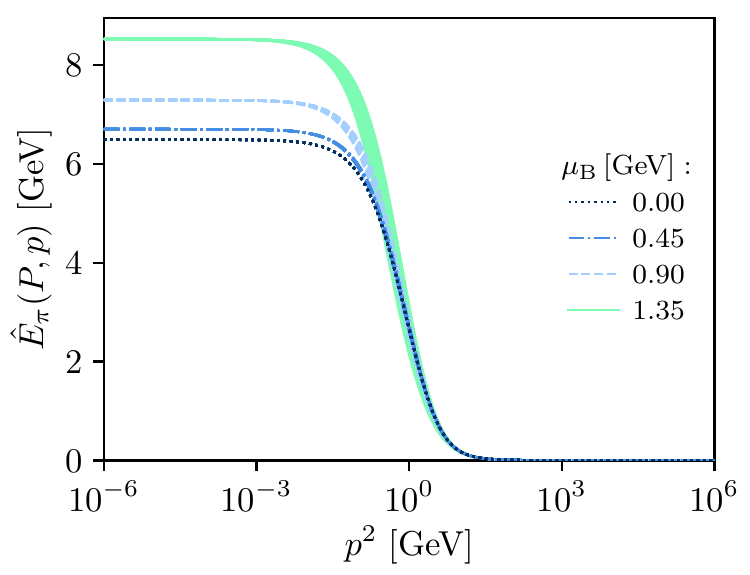}%
\vspace*{-3mm}%
\caption{\label{fig:quark_condensate_bsa_results}%
Left: Shifted light-quark condensate against the baryon chemical potential for the chirally-broken Nambu 
(solid black) and chirally-restored Wigner (dashed gray) solution. The boundaries for the
mixed region are denoted by $\muB^{\textup{N},\textup{W}}$ and indicated by vertical dotted lines. Right: First normalized on-shell pion BSA component $\hat{E}_{\pi}$ against the relative momentum $p^{2}$ between the quarks for various baryon chemical potentials. The spread of the amplitude corresponds to its dependence on the angle between $P$ and $p$. Only the Nambu solution is used for these results.
}
\end{figure}

To be able to distinguish between different phases of matter, we consider the quark condensate as the order parameter for chiral symmetry. The quark condensate for finite chemical potential and vanishing temperature is given by
\beq
\label{quark_condensate}
\expval{\bar{\Psi}\Psi}_f = -\Nc \+ Z_2^f Z_m^f \+ \int\frac{\dd^4 q}{(2\pi)^4} \Tr \bigl[ S_f(q) \bigr]
\eeq
with the quark mass and wave function renormalization constants $Z_m^f$ and $Z_2^f$, respectively. The light-quark condensate $\expval{\bar{\Psi}\Psi}_{\ell}$ (shifted by a constant) is plotted against the baryon chemical potential $\muB=3\+\muq$ on the left side of Fig.~\ref{fig:quark_condensate_bsa_results}. We determined the condensate for two types of stable  solutions of the quark DSE: the chirally-broken (Nambu) and the chirally-symmetric (Wigner) solution. The two solutions appear and disappear for certain chemical potentials and therefore mark three different regions in the plot. Above $\muB^{\textup{N}}$ the Nambu solution disappears and only the Wigner solution remains, while below $\muB^{\textup{W}}$ only the Nambu solution is found as an attractive solution. In between $\muB^{\textup{W}}$ and $\muB^{\textup{N}}$, both solutions can be found. The boundaries of this mixed region are given by 
\beq
\begin{aligned}
\muB^{\textup{W}}&=\SI{0.956}{\giga\eV} \, , \\
\muB^{\textup{N}}&=\SI{1.715}{\giga\eV} \, .
\end{aligned}
\eeq
The physical first-order phase transition, however, occurs in between these two chemical potentials and has to be determined from thermodynamical considerations. The light-quark condensate of the Nambu solution stays constant almost up to the end of the mixed region at $\muB^{\textup{N}}$. Since the nucleon mass is even smaller than $\muB^{\textup{W}}$ the condensate naturally satisfies the Silver-Blaze property\footnote{One has to bear in mind that strictly speaking the condensate is not an observable.}. The precision is even quite high: up to the chemical potential of the nucleon the shifted quark condensate only increases by $0.001\,\%$. Considering the Wigner solution, the quark condensate increases for lower chemical potentials and settles for higher ones.

\subsection{\label{sec:bsa_results}Bethe-Salpeter amplitudes}

On the right hand side of Fig.~\ref{fig:quark_condensate_bsa_results} we display the first normalized component of the pion BSA $E_\pi$ for finite chemical potential. We observe a strong increase of $E_\pi$ with chemical potential at small momenta, whereas the large momentum behavior is hardly affected. An even stronger chemical-potential dependence in the small- and mid-momentum regime can be found for the second pion BSA component $F_\pi$ (not shown in the plot). While the second BSA component of the sigma meson is similar to the one of the pion, the first BSA component is almost chemical-potential independent. We have furthermore demonstrated that the neutral pion is no longer invariant under charge conjugation at finite chemical potential \cite{Gunkel:2019xnh}. 

\subsection{\label{sec:meson_properties_results}Meson properties}

\begin{figure}[t]
\centering
\includegraphics[width=0.47\textwidth]{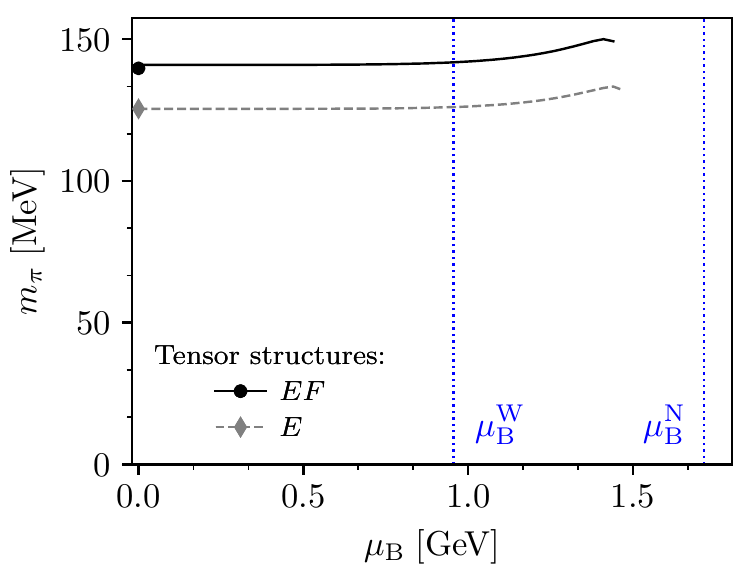}%
\hfill%
\includegraphics[width=0.47\textwidth]{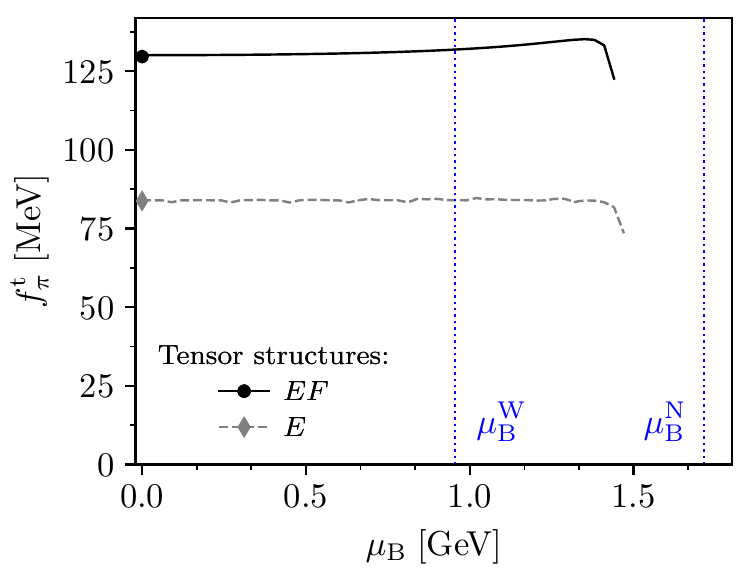}%
\vspace*{-3mm}%
\caption{\label{fig:pion_properties_finite_mu}%
Pion mass (left) and temporal pion decay constant (right) against the baryon 
chemical potential for different combinations of tensor structures used in the BSE calculation. 
The circle and diamond symbols represent the vacuum BSE calculations. All results are calculated with the
chirally-broken Nambu solution of the quark DSE.}
\end{figure}

We display the pion mass and temporal pion decay constant in vacuum and at finite chemical potential in Fig.~\ref{fig:pion_properties_finite_mu}. The spatial decay constant is not accessible since we consider the meson to be in the rest frame. In the plot, we see that the inclusion of the second BSA component $F_\pi$ into the calculation leads only to quantitative changes for the pion mass while the decay constant behavior changes qualitatively. Interestingly, including tensor structures beyond $E_\pi$ increases numerical stability. Compared to the corresponding values in vacuum, the pion mass increases by less than $1\,\%$ up to baryon 
chemical potential of the nucleon. The temporal decay constant increases by less than $2\,\%$. Thus within 
numerical precision both meson properties satisfy the Silver-Blaze property. For higher chemical potentials the variations increase up to a chemical potential where we were no longer able to obtain solutions. Whether this is connected to technical problems or with the first-order phase transition remains to be investigated. Our results for the sigma meson are discussed in Ref.~\cite{Gunkel:2019xnh}. For chemical potentials that 
are accessible in our calculation, the sigma meson satisfies the Silver-Blaze property as well. 

Our results are in line with the ones of previous works using functional methods \cite{Maris:1997eg,Bender:1997jf,Jiang:2008rb} where the BSE has not been 
solved explicitly but meson masses and decay constants have been extracted using Gell-Mann-Oakes-Renner type relations. Within the limits of these approximations, the Silver-Blaze property for meson observables 
has been observed at least approximately. Here we have solved an elaborated truncation scheme which takes 
into account the full momentum dependence of the gluon, determined the quark propagator in the complex momentum plane, and evaluated the corresponding BSEs numerically. Although the
quark propagator and the Bethe-Salpeter vertex function vary considerably with chemical potential, there
is a highly non-trivial cancellation mechanism in place, which renders the meson masses and decay constants independent of chemical potential and constitutes the Silver-Blaze property of QCD.   

\ack

This work has been supported by HGS-HIRe for FAIR, the GSI Helmholtzzentrum f\"{u}r Schwerionenforschung, the Helmholtz International Center for FAIR within the LOEWE program of the State of Hesse, and the BMBF under contract No.~05P18RGFCA.

\bibliographystyle{iopart-num}
\bibliography{mesons_finite_mu}

\end{document}